\documentclass[a4paper,UKenglish]{lipics}

\usepackage[dvipsnames]{xcolor}
\usepackage{amsmath}
\usepackage{amssymb}
\usepackage{amsthm}
\usepackage{stmaryrd}
\usepackage{mathpartir}
\usepackage{tikz}
\usepackage{enumitem}
\usepackage{bbding}
\usepackage{wasysym}
\usepackage[utf8]{inputenc}

\title{Easyprove: a tool for teaching precise reasoning}
\author[1]{Marek Materzok}
\affil[1]{Institute of Computer Science\\University of Wroc{\l}aw, Poland\\\texttt{marek.materzok@cs.uni.wroc.pl}}

\authorrunning{M. Materzok} %mandatory. First: Use abbreviated first/middle names. Second (only in severe cases): Use first author plus 'et. al.'

\Copyright{Marek Materzok}%mandatory. Default is "by";  http://creativecommons.org/licenses/by/3.0/

% TODO mandatory: Please choose ACM 1998 classifications from http://www.acm.org/about/class/ccs98-html . E.g., cite as "F.1.1 Models of Computation". 
% TODO mandatory: Please provide 1-5 keywords

\serieslogo{logo_ttl}%please provide filename (without suffix)
\volumeinfo%(easychair interface)
  {M. Antonia {Huertas}, Jo\~ao {Marcos}, Mar\'ia {Manzano}, Sophie {Pinchinat}, \\
  Fran\c{c}ois {Schwarzentruber}}% editors
  {5}% number of editors: 1, 2, ....
  {4th International Conference on Tools for Teaching Logic}% event
  {1}% volume
  {1}% issue
  {129}% starting page number
\EventShortName{TTL2015}
\begin{document}

\maketitle

\begin{abstract}
Teaching precise mathematical reasoning can be very hard.
It is very easy for a student to make a subtle mistake in a proof
which invalidates it, but it is often hard for the teacher
to pinpoint and explain the problem in the (often chaotically
written) student's proof.

We present Easyprove, an interactive proof assistant aimed
at first year computer science students and high school students,
intended as a supplementary tool for teaching logical reasoning.
The system is a Web application with a natural, mouse-oriented
user interface.
\end{abstract}

\section{Introduction}

A logic course is an essential part of every computer science curriculum.
Computer science students in University of Wrocław take a mandatory
course, called ``Logic for Computer Scientists'', during their first semester.
The main goal of the course is to teach the students the ability to
perform precise mathematical reasoning -- which is a very important skill for
theoreticians and practitioners alike. To achieve this goal, the students
are given a number of exercises which involve proving simple theorems, mostly
in the domain of set theory. 
The students are expected to provide informal proofs in natural language.
The emphasis is placed on logic as a tool for convincing yourself and others
of the validity of mathematical statements, and not on any particular logical formalism.

The students consider this course to be very hard. 
During the first few weeks, they struggle to understand the structure of a
mathematical proof.
The teachers do their best to guide them, but the time a teacher can give
to an individual student is necessarily very limited.
Therefore there is a strong need for an educational tool to aid
the learning process.
Such tool would serve as a playground, allowing the student to experiment
freely with writing mathematical proofs.
It should preferably both constrain the student, so that only correct 
proofs are accepted, and suggest him possible ways to continue the proof 
if he gets stuck.

There is a class of computer programs called proof assistants, which 
aid in formalizing proofs of mathematical theorems; examples
include Coq, HOL Light, Isabelle and Mizar.
These tools, while very useful for a specialist, are in our opinion
too complex and quirky to be used for teaching logic to freshmen.
Some have adapted them for use in  teaching
logic~\cite{Hendriks-al:10,Sakowicz-al:07,Halvorsen:MSc,Hoover-al:96};
we believe these tools are useful only for later stages of 
student's education.
There also already exist a number of tools developed specifically
as an aid for teaching logic, such as Jape~\cite{Bornat-al:97}, 
Yoda~\cite{Machin-al:11} and Panda~\cite{Gasquet-al:11},
but these tools focus mostly on teaching logic as a formalism
(natural deduction, sequent calculus, etc.), not as a method of
reasoning, as embodied by pen-and-paper proofs 
which can be found in mathematical journals.

We developed Easyprove as a supplementary teaching tool to be
used in our logic course for freshmen at the
University of Wrocław; we believe it might also find use in
high schools.
We briefly summarize our requirements in
the next section.

\section{Design goals}

In order to be most useful for teaching mathematical reasoning to 
freshmen, Easyprove was designed with following goals in mind:

\begin{description}[style=unboxed,leftmargin=0px,itemsep=1em]
\item[Easy to access.]
A good teaching tool should be easy to access for the student
both during classes and at home. 
Using contemporary IT technology, the best way to achieve
this is to design the tool as a web application. 
The Javascript engines of today's Web browsers allow to
create Web interfaces which are just as usable as classical ``desktop'' ones.
One can also argue that young people are now very familiar with
Web-like interfaces thanks to the popularity of smartphones
and tablets.

The approach of designing an educational tool as a Web app
was successfully applied by many other projects, including
Yoda~\cite{Machin-al:11}, BoxProvr~\cite{Halvorsen:MSc}
and ProofWeb~\cite{Hendriks-al:10}.

\item[Natural syntax.]
To use a full-featured proof assistant, the user has to learn a specialized
syntax used for representing terms, proofs and commands for the
system. 
This can be a big burden for a computer science freshman, 
who is not yet familiar with any programming language or other
formal syntax. 

To remedy this problem, Easyprove presents proofs using 
a notation close to natural
language, and the terms are displayed using Unicode mathematical symbols,
which are already familiar to the student from mathematics classes.
The symbols can be entered using a graphical keyboard,
LaTeX-like shorthands, or copied-and-pasted from other sources.

\item[Familiar setting.]
Many proof assistants are based on complex formalisms, which
can be hard to understand to freshmen. For example, HOL and Coq
use higher-order logics, and it is typical to encode sets as
predicates within the logic. This is often advantageous for an
expert user, but for a beginner it is causing many unexpected
problems: e.g., simple terms like $\{\{a\}\} \cup \{a\}$,
which the student encounters on mathematics classes, give
a type error.

In contrast, Easyprove is based on first-order logic and 
(Zermelo-Fraenkel) set theory.
First-year students should already have some basic familiarity 
with them from mathematics classes.

\item[Discoverable user interface.]
Proof assistants have in general a very steep learning curve. 
Some of them present a command prompt to
the user, who is required to learn a big number of commands or
``tactics'' in order to use them (e.g. Coq, HOL Light). 
Others work
in batch mode -- the user has to write a proof script, which is then
checked for correctness by the tool (e.g. Mizar, Isabelle). 
% TODO is there a command-line interface for Isabelle (not Isar)?
There are GUI front-ends available, but these are of little help
for learning: they usually list all possible commands in menus,
the commands are not well described, and most of them are not
applicable in a given context.

In contrast, Easyprove presents the user with applicable
actions only, and their effects are completely described.
Also, every function of Easyprove is accessible by the graphical
interface, which can be used with only a mouse.

\item[Features for classroom use.] Easyprove is designed with
features important for teachers in mind: creating task lists,
managing student accounts, storing students' solutions on the 
server for review and replay. It also supports internationalization:
we have implemented Polish and English front-end language versions.
\end{description}

In the next section we present how Easyprove meets these
design goals by presenting example interactions with the
system.

\section{Using Easyprove}

To run Easyprove, one enters its URL in any Web browser with standard
CSS and Javascript support. 
One is then presented with a screen with
a list of pre-entered tasks -- theorems to prove. 
One can also create a new task from scratch.
To create a task, the user has to give it a name, enter the goal formula,
and select the set of assumptions (axioms or lemmas) which
will be available in the task.

\subsection{Term editor}

\begin{figure}
\begin{center}
\includegraphics[width=0.7\textwidth]{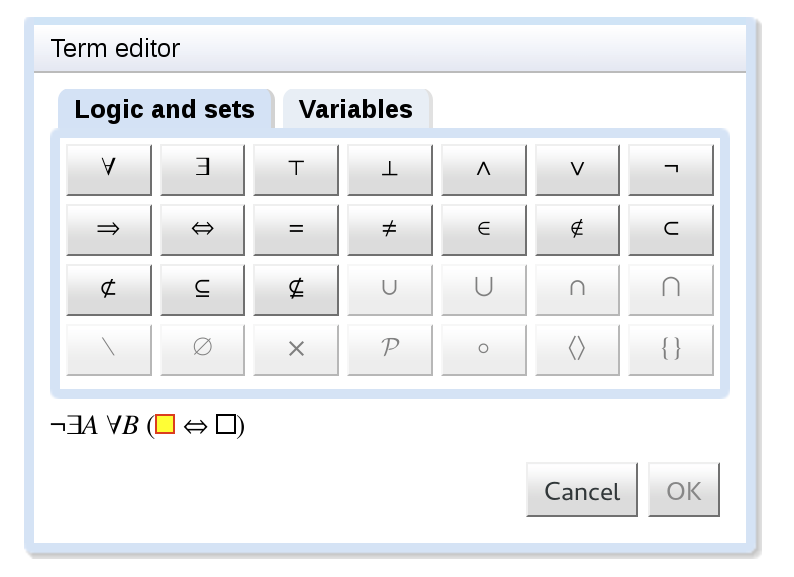}
\end{center}
\caption{Term editor}
\label{fig:termeditor}
\end{figure}

The goal is entered using a graphical term editor (Figure~\ref{fig:termeditor}).
The editor window has two main parts: on the top is the visual keyboard,
which presents available symbols and variable names present in
the current context; the currently entered term is displayed below it. 
In the visual keyboard,
each button has a tooltip which displays a short description
and presents how a given symbol can be entered using a keyboard.

The editor has two modes of operation: structural and linear.
In the structural mode, clicking a button on the visual keyboard
fills the currently selected hole in the term, possibly creating
new holes for the subterms. When the user starts typing or clicks
on a hole, the editor switches to the linear mode. The visual keyboard
can now be used to insert a symbol in the current cursor position.
Symbols can also be entered by typing a LaTeX-like shorthand,
e.g., typing \texttt{{\textbackslash}forall} causes the symbol $\forall$ to appear.
Pushing the Enter button parses the entered term and switches
the editor back to visual mode. If a mistake is made, clicking on
a subterm returns it to linear mode for editing.

\subsection{Proof editor}

\begin{figure}[t]
\begin{center}
\includegraphics[width=0.99\textwidth]{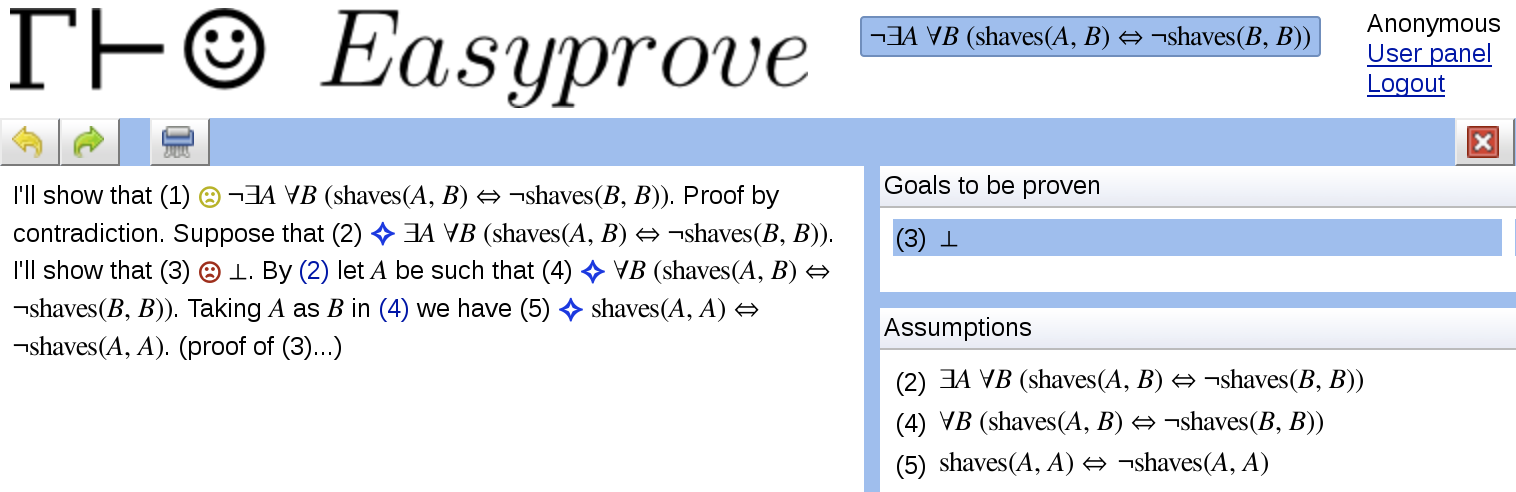}
\end{center}
\caption{Proof editor}
\label{fig:proofeditor}
\end{figure}

The main part of Easyprove is the proof editor. 
The proof editor screen (Figure~\ref{fig:proofeditor}) consists of
the main block, the sidebar and the toolbar. 
The main block presents the proof in a form resembling a pen-and-paper
mathematical proof. 
The sidebar lists the current goals and the assumptions applicable for the
currently selected goal, which is highlighted.
There can be many goals active, one for each branch of the proof 
created by case analysis or unproved lemmas; the user is free to
switch between them at any time.
The toolbar contains the undo/redo buttons
and a ``delete last step in this proof branch'' button. 

The formulas occurring in the proof text are of one of two kinds:
they can be either a goal or an assumption. The currently applicable
assumptions are marked with a blue star~{\color{blue} \scriptsize \FourStarOpen};
the ones from other proof branches are marked with a gray star.
Unproved goals are marked with a red sad face {\color{red} \frownie};
when the goal is not active, but its proof branch is not yet
proved completely, the face turns yellow~{\color{YellowOrange} \frownie}.
A proved goal is distinguished by a green happy 
face~{\color{ForestGreen} \smiley}.
Every formula in the proof text is numbered, so that it can be
referred to using this number both in the proof text and in the
sidebar.

Initially the proof has only one goal -- the theorem one wants
to prove -- and no assumptions (the assumptions selected when creating
a task are implicit and are not shown in the proof).
To illustrate how one writes a proof in Easyprove,
we use the classical barber paradox as an example.
In particular, we will prove that the existence of
a barber which shaves every person that does not shave himself
is paradoxical. 
This can be stated in first order logic as:\footnote{This sentence
can be directly copied from this article's PDF file and pasted
into the term editor.}
$$
\neg\exists A \forall B (\textrm{shaves}(A,B) \Leftrightarrow \neg \textrm{shaves}(B,B))
$$
Interaction with the proof editor is inspired by
proof by pointing~\cite{Bertot-al:94}.
Clicking on a subterm of the goal (the subterm currently
pointed at is highlighted) causes a menu window to appear,
which lists proof steps applicable to the selected subterm.
In our example, clicking on the whole term opens the window
presented below:
\begin{center}
\includegraphics[scale=0.3]{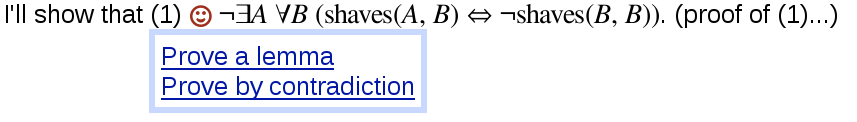}
\end{center}
The proof step titled ``prove a lemma'' is always available, and
allows to introduce lemmas. The next one, ``prove by contradiction'',
is also always available, and is the one we want to use here. 
When the user hovers the mouse over a proposed proof step, a tooltip
appears:
\begin{center}
\includegraphics[scale=0.3]{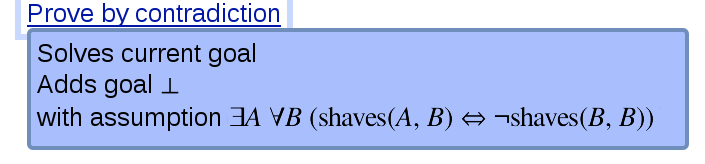}
\end{center}
The tooltip briefly summarizes how the proof will be extended
if the user activates the selected proof step. In this case, the
negated goal will be added as an assumption, and the new goal will
be $\bot$ -- a contradiction.

Clicking on the new assumption will show a proof step named
``take this'', which corresponds to existential quantifier elimination.
Selecting this step will add an assumption about the variable
$A$ -- which symbolizes the paradoxical barber:
\begin{center}
\includegraphics[scale=0.3]{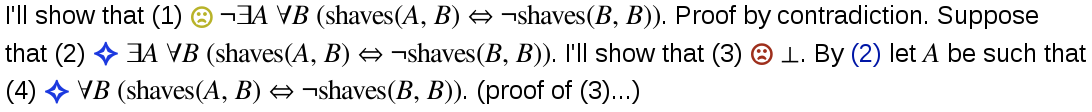}
\end{center}

The new assumption (4) is an universal sentence, so a proof step
named ``specialize'' is available for it. Choosing it invokes the
term editor, which allows to enter the term used to specialize
the universal sentence. Here we intend to examine the case when
the person $B$ is the barber $A$ himself, so we choose $A$ from
the ``Variables'' tab (the sentence shown next to the button is
the first assumption mentioning this variable):
\begin{center}
\includegraphics[scale=0.3]{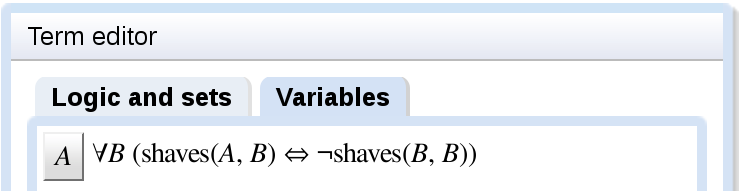}
\end{center}

We get the obviously contradictory sentence
$\textrm{shaves}(A,A) \Leftrightarrow \neg \textrm{shaves}(A,A)$
(the proof state at this point is shown in Figure~\ref{fig:proofeditor}).
Contradiction can be derived from it either by using 
a built-in lemma about equivalence, or by case analysis using
the law of excluded middle; we show below the completed proof using the first
possibility:
\begin{center}
\includegraphics[scale=0.3]{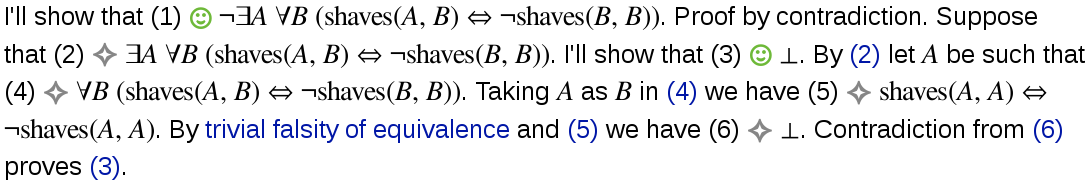}
\end{center}

\subsection{Proving theorems in set theory}
\label{sec:settheory}

\begin{figure}[t]
\begin{center}
\includegraphics[width=0.99\textwidth]{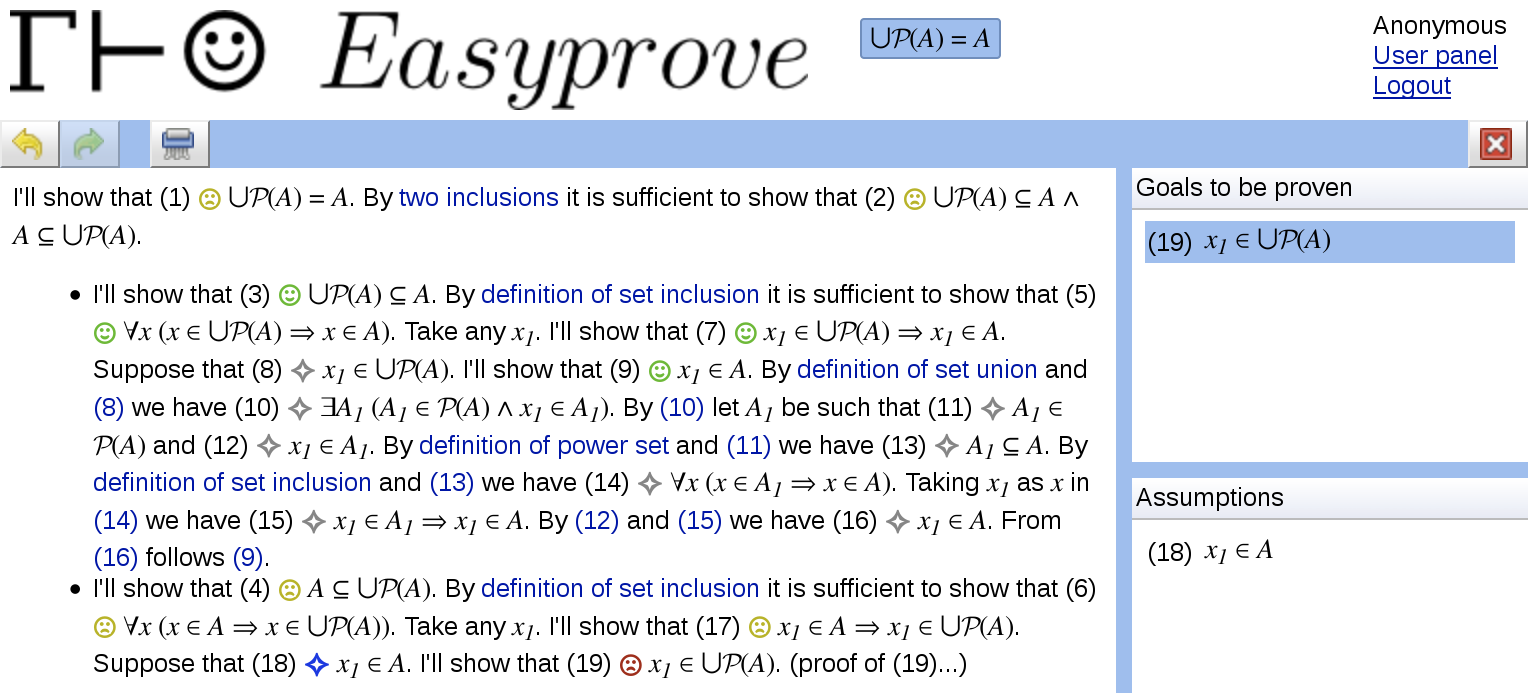}
\end{center}
\caption{Proof editor -- proof in set theory}
\label{fig:proofeditorset}
\end{figure}

Easyprove was specifically designed for writing proofs of simple
theorems in set theory. To illustrate its features, we now
briefly present a proof that for every set $A$, the union of all its
subsets is equal to $A$; i.e. that $\bigcup(\mathcal{P}(A))=A$.

Set equality is usually proven using the extensionality principle,
or equivalently by proving two inclusions. Both principles are
available as lemmas in Easyprove, here we choose the second one:
\begin{center}
\includegraphics[scale=0.3]{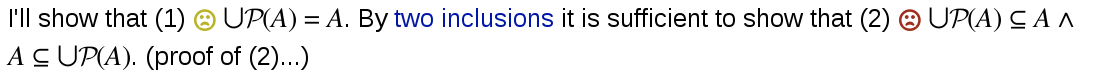}
\end{center}

We get a goal which is a conjunction, so we use a proof step named
``prove conjuncts'' to prove the conjuncts separately. We can then
apply the definition of set inclusion to both goals (the option
to do so is present in the drop-down menus for the goals):
\begin{center}
\includegraphics[scale=0.3]{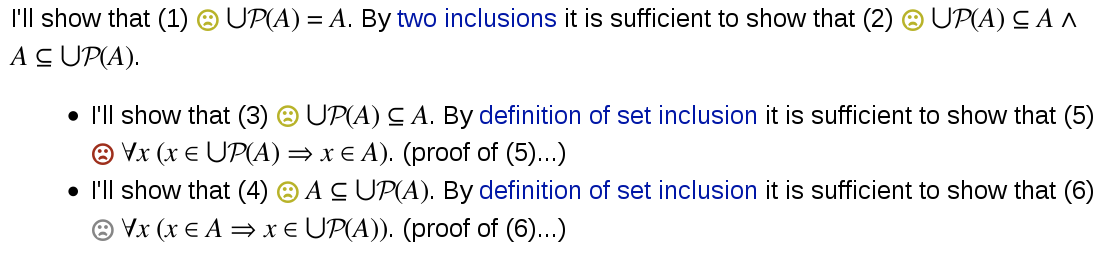}
\end{center}

The proof can easily be completed in this way, by repeatedly
applying proof steps from logic and using set-theoretic definitions
and lemmas. A partially completed proof of the theorem in Easyprove
is shown in Figure~\ref{fig:proofeditorset}.

\section{Implementation}

Easyprove is mostly implemented using the programming language Java.
The Google Web Toolkit\footnote{\url{http://www.gwtproject.org}}
is used for the user interface, for communicating with the Easyprove
server, and for compiling Java code to Javascript, which can be
executed in a browser. 
Most of the Easyprove code, including the
proof engine, runs on the client side, which reduces the load placed
on the server and allows easy scaling for large number of users.
The Easyprove server, which manages the task list and user accounts,
is implemented in Scala\footnote{\url{http://www.scala-lang.org}}, and runs on any Java Servlet container
(e.g. Tomcat, Jetty). 

The proofs are represented internally as trees, with nodes 
corresponding to each proof step used by the user. The nodes
may be one of two kinds: they may represent forward reasoning
steps, which add new assumptions, but do not change the goal;
or they can change the current goal or add new goals.
Either way, discarding assumptions (weakening) is prohibited 
-- an introduced assumption is valid for its entire proof branch. 
This corresponds
to the way assumptions are usually treated in pen-and-paper
proofs. The reasoning rules implemented in the Easyprove system are
sound and complete with respect to the classical first-order logic.

The new formulas introduced by proof steps (assumptions or goals)
are automatically simplified according to the following rules: 
\begin{itemize}
\item Expressions with binary logical operators 
\item Truth $\top$ and falsity $\bot$ symbols are propagated 
(e.g. $\phi \vee \top$ becomes $\top$) or eliminated
($\phi \wedge \top$ becomes $\phi$) if possible;
$\vee$ and $\wedge$ are reassociated to the left;
\item Negations are pushed inwards using the de Morgan's laws,
double negations are eliminated.
\end{itemize}
The justification is that a mathematician writing a proof
is not concerned with the particular shape of the formulas
he is working with, but with their meaning. 

Assumptions applicable to a clicked subterm on some goal
or assumption (see Section~\ref{sec:settheory}) are found
using pattern matching. For example, the definition of
set inclusion is stored in Easyprove as:
$$\forall A \forall B (A \subseteq B \Leftrightarrow \forall x (x \in A \Rightarrow x \in B))$$
The system finds this assumption to be applicable to the
goal $\bigcup(\mathcal{P}(A))\subseteq A$ by matching it to
the pattern $A \subseteq B$, where $A$ and $B$ are considered
pattern variables. To make the search for matching assumptions
efficient when the number of them is large, the patterns
are stored in a prefix tree (trie) structure, where the keys
are pre-order representations of the patterns -- e.g.
$A \subseteq B$ is stored in the prefix tree as ``$\subseteq,A,B$''.

\section{Conclusions and future work}

We have implemented Easyprove, an easy to use proof assistant,
as an aid for teaching logical reasoning. 
It is targeted for first year computer science students and 
high school students.
The system is currently in prototype stage and requires further
work in order to be fully usable.
The goals for future development are:

\begin{description}[style=unboxed,leftmargin=0px,itemsep=1em]
%\item[Extension of the logic engine.] 
%Some proofs are currently very awkward to write in Easyprove.
%A lot of important syntactic sugar is missing, including 
%defining finite sets by listing their elements (e.g. $\{a,b,c\}$),
%restricted quantification ($\forall X \in A \  \phi(x)$),
%set comprehension with patterns ($\{\langle a,b \rangle \in A \times B \ |\  \phi(a,b) \}$).
%There is no support for arithmetic and induction, excluding many
%interesting tasks to be expressed in Easyprove.
\item[Other axiom systems.]
Easyprove can be in principle extended to handle axiomatic
systems other than set theory -- for example, Hilbert's
axioms for Euclidean geometry. Such an extension would
enhance the educational utility of the tool.
\item[Better proof editor.] 
The proof editor does not currently allow easy modification
of the proof: the only way to restructure an already entered
part of the proof is to undo the most recent proof step
in a proof branch. 
A user interface for restructuring a proof would aid active
experimentation, and therefore improve the learning experience.
\item[More use of natural language.] 
Easyprove currently presents formulas in symbolic form. 
Adding an option to present them in natural language instead
(e.g. instead of $\forall x \  \textrm{shaves}(x,y)$ write
``for all $x$, $x$ shaves $y$'') would help the student make
the connection between what he sees on screen and actual
pen-and-paper mathematical proofs.
\item[Proof automation.] 
The main priority in Easyprove's design was handling short
proofs of simple theorems well, because these are most instructive
for teaching logical reasoning. Because of this decision,
writing longer proofs in Easyprove is tedious. It remains
to be explored how one can include automation in Easyprove
without hurting its didactic character.
\item[Mobile interface.] 
Easyprove was designed with a traditional keyboard and mouse
in mind. Given the recent popularity of touch screen devices,
a redesign of the user interface to accommodate them would certainly
allow Easyprove to reach a wider audience.
\end{description}
Nevertheless, the system has proven to be usable in its current
state. It was not used in the classroom yet, but it was presented
to selected students and their response was positive. 

The readers are invited to try out Easyprove. The current stable
version of the system is available online at
\url{http://easyprove.ii.uni.wroc.pl/}.

\bibliographystyle{plain}
\bibliography{logbib}

\end{document}